\documentclass[12pt,a4]{article}
\usepackage{latexsym}
\usepackage{amsmath}
\usepackage{bm}
\usepackage{graphicx,color}
\usepackage{txfonts}
\newcommand{\bea}{\begin{eqnarray}}
\newcommand{\eea}{\end{eqnarray}}
\newcommand{\nn}{\nonumber}

\title
{Bargmann-Wigner Equations, Fermion-Boson Correspondence and Superradiant Problem 
in Curved Spacetime }
\author{
Masakatsu KENMOKU\thanks{m.kenmoku@cc.nara-wu.ac.jp} \\
Nara Women's University, Kitauoya, Nara 630-8306, Japan \\
Nara Science Academy, 178-2 Takabatake, Nara 630-8301, Japan
}
\date{\empty}

\begin{document}
%====================================================
\maketitle
\abstract{
Bargmann-Wigner equations and their solutions are studied in (3+1)-dimensional curved   spacetime. 
Fermion-Boson correspondence for bi-spinor case is studied 
through the Bargmann-Wigner equations and solutions over curved spacetime. 
As an application to scattering phenomena of massive Fermions and Bosons 
on the rotating black holes,  
the superradiance with negative energy $(\omega<0)$ 
and positive effective energy in co-rotating coordinate system near horizon 
$(\omega-m\Omega_{H}>0)$ is possible to occur 
as stable physical states in Kerr spacetime. 
}

%------------------------------------------
\section{Introduction}
%-------------------------------------------
\renewcommand{\theequation}{\thesection.\arabic{equation}}
\setcounter{equation}{0}

%%%%%%%%%%% motivation %%%%%%%%%%%%%%%%%%%%%%%

It is important to make clear the matter field dynamics in a curved spacetime  
for the astrophysical observation and theory.  
As matter fields, Fermions and Bosons have different features in one side 
and have mutually related features in other side.  
One of the important theoretical observation for scattering of matter fields 
is superradiant problem of Fermion and Boson fields 
in the rotating black hole geometry. 
The superrsiant phenomena were studied not to occur for Fermion fields 
\cite{Unruh1974, Maeda1976, Wagh1985}
but to occur for Boson fields 
 \cite{Unruh1974,Press1972,Misner1972,Detweiler1980,Takasugi1997, Mukohyama2000}
in (3+1)-dimensional rotating black hole spacetime. 
The successive occurrences of superradiant phenomena for Boson fields were studied to  
cause the serious problem of the instability of black holes especially in Kerr geometry  
\cite{Press1972, Chandrasekhar1983,Cardoso2004,Kodama2008}. 
Therefore the Fermion and Boson relation is one of the important theoretical key to solve the 
cosmological problems including the superradiant problem. 

%%%%%%%%%%%%%%%%%%%%%%%%%%%%%%%%%%%%%%%%%%%%%%%%%%%%%%%%%%
As a kinematical relation between Fermions and Bosons, 
the field equations for 
the direct product of spinor fields are studied to derive those for the higher spin states 
originally by Bargmann and Wigner in flat Minkowski spacetime,  
which is known as the Bargmann-Wigner (BW) equations \cite{Bargmann1948,Lurie}.
The BW formulation is expected to analyze the superradiant problem as one of effective applications. 
 
%%%%%%%%%%%%% purpose of this paper %%%%%%%%%%%%%%%%%%%%%%%%
In this paper, we extend the BW equations in flat Minkowski spacetime to 
those in general curved spacetime in (3+1)-dimensions. 
Explicitly we study the BW equations for bi-spinor fields 
as the direct product of two spinor fields to obtain 
the Boson field equations for pseudoscalar and vector fields. 
In this approach, we obtain the consistent solutions for Fermion fields in terms of Boson fields 
under the torsion free condition. 

%%%%%%%%%%%%%%%%%%%% results %%%%%%%%%%%%%%%%%%%%%%%%%%%%%%%
Applying the result to the superradiant problem,  
we obtained the result that spinor fields as Fermi particles 
and pseudoscalar and vector fields as Boson particles should 
show the corresponding behavior in scattering problem in curved spacetime 
because they are related by the BW equations.  
This means that both of them occur superradiant instability or both do not in rotating black hole geometry. 
    
In addition to this result, 
we will study the type of the superradiance; stable or unstable, 
considering the co-rotating coordinate frame in Kerr spacetime 
and obtain the spectrum condition 
that the effective energy in co-rotating coordinate system 
near the outer horizon of black hoke is positive:  $ \omega-m\Omega_{H}>0$ . 
This result is compared with the general superradiant condition 
$\omega(\omega-m\Omega_{H})<0$, 
the superradiance occur as $\omega-m\Omega_{H}>0$ with $\omega<0$ 
for both Fermions and Bosons.  We call this as type 2 superradiance. 
The result is consistent with our previous result in which the type 2 superradiant phenomena 
are studied as stable physical normalizable states.   
The negative particle energy  ($\omega<0$) is necessary because of the completeness relation of matter fields 
and the type 1 superradiance ($\omega-m\Omega_{H}>0$ with $\omega<0$) is not normalizable physical states. 

%%%%%%%%%%%%%%%%%%%%%% comment %%%%%%%%%%%%%%%%%%%%%%%%%%%%%%%%%%%%%%%

We note that in our previous paper, we derived current relations between spinor and scalar fields 
to discuss the superradiant problem 
in asymptotic flat spacetime region of Kerr geometry using the BW formulation \cite{KC2015}. 
In this paper we  derive the direct relation of field equations and solutions 
between Fermi and Bose particles  
in full spacetime region of general curved geometry.  
We also note that, in this formulation, the mass of matter fields should be finite   
because Fermion and Boson fields have different mass dimensions.   

It is very interesting to note that, as one of massive pseudoscalar fields, 
axion fields is important in the superradiant problem and one of the dark matter candidates 
\cite{Yoshino2012,Yoshino2015}.    
 
%%%%%%%%%%%%%%%%%%%%%% organization of this paper %%%%%%%%%%%%%%%%%%%%%%%%%%%

The organization of this paper is as follows.
In section 2, the BW equations for bi-spinor 
in flat Minkowski spacetime are extended to include 
scalar part as well as vector parts in (3+1)-dimensional general curved spacedtime. 
In section 3, Fermion and Boson correspondence is studied using the BW equations and solutions. 
In section 4, the superradiant problem is studied in the co-rotating coordinate frame 
in general rotating spacetimes including Kerr geometry. 
Summary and discussion will be given in the final section.

%%%%%%%%%%%%%%%%%%%%%%%%%%%%%%%%%%%%%%%%%%%%%%%%%%%%%%%%%%%%%%%%%
%%%%%%%%%%%%%%%%%%%%%%%%%%%%%%%%%%%%%%%%%%%%%%%%%%%%%%%%%%%%%%%%%

\section{Bargmann-Wigner equations in curved spacetime}
\setcounter{equation}{0}
In this section, we derive the BW equations in general curved spacetime 
to describe the Boson states consistently with the fundamental spinor states. 
For this purpose, we start to consider the BW equations for bi-spinor fields 
in a local Minkowski spacetime attached to each point of the general curved spacetime.  

%%%%%%%%%%%%%%%%%%%%%%%%%%%%%%%%%%%%%%%%%%%%%%%%%%%%%%%%%%%%%%%%%%%%
\subsection{Bargmann-Wigner equations in local Minkowski spacetime}
%%%%%%%%%%%%%%%%%%%%%%%%%%%%%%%%%%%%%%%%%%%%%%%%%%%%%%%%%%%%%%%%%%%%
The B-W equations for the bi-spinor field $\Psi(x)^{(\rm BW)}$ are written  
\cite{Bargmann1948,Lurie}: 
\begin{eqnarray}
(\gamma^{a} {\partial}_{a}+M)_{\alpha\alpha'} {\Psi(x)}^{(\rm BW)}_{\alpha'\beta}&=&0 ,  \nonumber  \\ 
(\gamma^{a} {\partial}_{a}+M)_{\beta\beta'} {\Psi(x)}^{(\rm BW)}_{\alpha\beta'}&=&0 ,  \label{BW01}
\end{eqnarray}
where $\gamma^{a}$ and $M$ denote the Dirac gamma matrices in a local Minkowski spacetime   
and a mass of the bi-spinor field.  The index $a=0,1,2,3$  labels the vector component in local Minkowski coordinate system.  
We rewrite the BW equations (\ref{BW01}) omitting spinor indexes $\alpha, \beta$ as    
\begin{eqnarray}
(\gamma^{a} {\partial}_{a}+M) {\Psi(x)}^{(\rm BW)}&=&0 ,  \nonumber  \\ 
{\Psi(x)}^{(\rm BW)}(\overleftarrow{{\partial}}_{a}\gamma^{a\ T}+M) &=&0
 ., \label{BW02} 
\end{eqnarray}
In order to avoid the transpose operation $T$ in eq. ($\ref{BW02}$), 
we introduce the modified bi-spinor BW field defined as 
\begin{eqnarray}
\Psi(x)=\Psi(x)^{(\rm BW)}C^{-1} ,
\end{eqnarray}
where $C$ denotes the charge conjugation operator and the relation $C\gamma^{a\ T}C^{-1}=-\gamma^{a}$ is used. 
Then we obtain the BW equations for the bi-spinor field $\Psi$ as
\begin{eqnarray}
(\gamma^{a}\partial_{a}+M)\Psi(x)&=&0 \ , \nonumber\\ 
\Psi(x) (\overleftarrow{\partial}_{a}\gamma^{a}-M)&=&0 \ . \label{BWinM}
\end{eqnarray}

%%%%%%%%%%%%%%%%%%%%%%%%%%%%%%%%%%%%%%%%%%%%%%%%%
\subsection{Relation between local Minkowski and curved spacetimes}

In general curved spacetime, the gamma matrices with Roman indexes ($a,b$)   
and their algebra are defined in local Minkowski spacetime
\cite{Bargmann1948,Weinberg2000}: 
\begin{eqnarray}
\{\gamma^{a}, \ \gamma^{b}\}=2\ \eta^{ab}\ , \
\eta^{ab}={\rm diag}(-1,1,1,1)\ . 
\end{eqnarray}
The gamma matrices with Greek indexes ($\mu, \nu$) and their algebra in general curved spacetime are obtained 
through the local Minkowski spacetime:   
\begin{eqnarray}
\gamma^{\mu}&:=&e_{a}^{\mu}\gamma^{a}\ , \nonumber \\
\{\gamma^{\mu}, \ \gamma^{\nu}\}&=&2\ g^{\mu\nu}\ , \
g^{\mu\nu}=e^{a\ \mu}\ e_{a}\ ^{\nu} \ ,
\end{eqnarray}
where $e_{a}^{\mu}$ and $g^{\mu\nu}$ denote a vierbein (or tetrad)  and the metric tensor.   

%%%%%%%%%%%%%%%%%%%%%%%%%%%%%%%%%%%%%%%%%%%%%%%

For spinor fields $\psi(x)$, they transform as scalars under general coordinate transformations but 
as spinors under a local Lorentz transformation $\Lambda(x)$  
\begin{eqnarray}
\psi(x) \rightarrow D(\Lambda(x))\psi(x) \ ,
\end{eqnarray}
where $D(\Lambda)$ is the spinor representation of the homogeneous Lorentz 
group. 
A covariant derivative for spinors is introduced as
\begin{eqnarray}
\mathcal{D}_{\mu}\psi=(\partial_{\mu}+\Omega_{\mu})\psi \ , 
\end{eqnarray}
which is defined to transform under local Lorentz transformations like $\psi$ itself: 
$\mathcal{D}_{\mu}\psi(x) \rightarrow D(\Lambda(x))\mathcal{D}_{\mu}\psi(x)\ .$ 
The connection matrix $\Omega_{\mu}$ for spinors is written as  
\begin{eqnarray}
\Omega_{\mu}= \frac{1}{4}\omega^{ab}_{\ , \mu}\Sigma_{ab} \ , \label{Omega}
\end{eqnarray}
where $\omega^{ab}_{\ ,  \mu}$ is the spin connection and 
$\Sigma_{ab}=(\gamma_{a}\gamma_{b}-\gamma_{b}\gamma_{a})/2$ are the spin matrices representing 
the generators of homogeneous Lorentz group. 
Then the covariant Dirac equation in general spacetime is derived as
\begin{eqnarray}
(\gamma^{\mu}\mathcal{D}_{\mu}+M)\psi(x)=0\ ,
\end{eqnarray}
with the mass parameter  $M$. 
%%%%%%%%%%%%%%%%%%%%%%%%%%%%%%%%%%%%%%%%%%%%%%%%%%%%%%%%%%%%%%%%%%%%%%%

For vector fields $A_{\nu}(x)$, they  transform as vectors under general coordinate transformations 
and their covariant derivative is defined as
\begin{eqnarray}
\nabla_{\mu}A_{\nu}
=\partial_{\mu}A_{\nu}-\Gamma^{\lambda}_{\nu\mu}A_{\lambda}\ , 
\end{eqnarray}
where $\Gamma^{\lambda}_{\ \ \nu\mu}$ denote the affine connection. 

The covariant derivatives for vierbeins with indexes of local Minkowski and general coordinates  
are expressed as  
\begin{eqnarray}
\mathcal{D}_{\mu}e^{a}_{\ \nu}
=\partial_{\mu}e^{a}_{\ \nu}+\omega^{a}_{\ b, \ \mu}e^{b}_{\ \nu}
-\Gamma^{\lambda}_{\mu \nu}e^{a}_{\lambda} \ .
\end{eqnarray}
The vierbein condition is imposed to determine the geometry: 
\begin{eqnarray}
\mathcal{D}_{\mu}e^{a}_{\ \nu}=0 \ , \label{condition}
\end{eqnarray}
which leads the explicit form of the spin connection 
\begin{eqnarray}
\omega^{ab}_{,\, \mu}&=&g^{\nu\lambda}e^{a}_{\nu} \nabla_{\mu}e^{b}_{\lambda} .
\label{omega}
\end{eqnarray}

%%%%%%%%%%%%%%%%%%%%%%%%%%%%%%%%%%%%%%%%%%%%%%%
\subsection{Bargmann-Wigner equations for bi-spinor and Boson fields in curved spacetime}

Next we consider the bi-spinor BW fields $\Psi(x)$ as possible Boson states.  
The bi-spinor fields transform under local Lorentz transformations 
as direct product of spinors from lefthand and righthand sides as 
\begin{eqnarray}
\Psi(x) \rightarrow D(\Lambda(x))\Psi(x) D^{-1}(\Lambda(x)) \ ,
\end{eqnarray}
which define the covariant derivatives as 
\begin{eqnarray}
\mathcal{D}_{\mu}\Psi(x):=
\partial_{\mu}\Psi+\Omega_{\mu}\Psi-\Psi\Omega_{\mu}\ , 
\end{eqnarray}
where we take into account transpose symmetry relations: 
$C\Lambda^{T}C^{-1}=-\Lambda$ and $C\Sigma_{ab}^{T}C^{-1}=-\Sigma_{ab}$.    
Then the covariant field equations for bi-spinors are derived
\begin{eqnarray}
(\gamma^{\mu}\mathcal{D}_{\mu}+M)\Psi(x) &=&0 \ , \label{BW1}\\
\Psi(x)(\overleftarrow{\mathcal{D}}_{\mu}\gamma^{\mu}-M)&=&0 \ .\label{BW2}
\end{eqnarray}
We expand the bi-spinor field in a set of Boson fields as 
\begin{eqnarray}
\Psi(x)=
aS(x)+b\gamma_{5}P(x)+c\gamma^{\mu}V_{\mu}(x)+d\gamma_{5}\gamma^{\mu}A_{\mu}(x))
+\frac{e}{2}\Sigma^{\mu\nu}F_{\mu\nu}(x) \ , \label{biB}
\end{eqnarray}
where $S, P, V_{\mu}, A_{\mu}$ and $F_{\mu\nu}$ denote scalar, 
pseudoscalar, vector, axial vector  and tensor fields respectively. 
Coefficients $a, b, c, d, e$ are determined to satisfy the BW equations (\ref{BW1}) and (\ref{BW2}). 

%%%%%%%%%%%%%%%%%%%%%%%%%% subtraction %%%%%%%%%%%%%%%%%%%%%%%%%%%%%%%%%%%%%%%%%%%%
Subtracting these equations: ($\ref{BW1})-(\ref{BW2}$), we find the set of relations among Boson fields:  
\begin{eqnarray}
aM S(x)&=& 0   , \ \label{S}\\
bM\gamma_{5}P(x) + d\gamma_{5} \nabla^{\mu}A_{\mu}(x)&=&0 ,  \ \label{PA} \\
cM \gamma^{\mu}  V_{\mu}(x)-e \gamma^{\mu}\nabla^{\nu} (F_{\mu\nu}(x)-F_{\nu\mu}(x))
&=&0 , \label{VF}\\
\frac{e}{2}M \Sigma^{\mu\nu}F_{\mu\nu}(x) +c \Sigma^{\mu\nu} \nabla_{\mu}V_{\nu}(x) &=&0  \ \label{FV} .
\end{eqnarray}
Coefficient parameters are chosen to adjust the field mass dimensions:   
\bea
a=0, b=c=d=M^2 , e=-M , 
\eea 
and independent field equations for Boson field relations are obtained from eqs. (\ref{S})-(\ref{FV}): 
\begin{eqnarray}
\nabla^{\mu}\partial_{\mu}P(x)-M^{2}P(x)&=& 0 , \label{Peq} \label{Peq}\\ 
\nabla^{\mu}(\nabla_{\mu}V_{\nu}(x)-\nabla_{\nu}V_{\mu}(x))-M^{2}V_{\nu}(x)&=&0 . \label{Veq}
\end{eqnarray}
Other Boson fields are given by independent fields as 
\begin{eqnarray}
S(x)&=&0 , \label{Seq} \\
MA_{\mu}(x)&=& \partial_{\mu}P(x) , \label{Aeq} \\
F_{\mu\nu}(x)&=&\nabla_{\mu}V_{\nu}(x)-\nabla_{\nu}V_{\mu}(x) . \label{Feq} 
\end{eqnarray}
 
%%%%%%%%%%%%%%%%%%%%%% summation %%%%%%%%%%%%%%%%%%%%%%%%%%%%%%%%%%%%%%%%%%%%%%%%%%
The sum of the BW eqs. (\ref{BW1})+(\ref{BW2}) is calculated to be 
\begin{eqnarray}
&&(\gamma^{\mu}\mathcal{D}_{\mu}+M)\Psi(x)
+\Psi(x)({\overleftarrow{\mathcal{D}}}_{\mu}\gamma^{\mu}-M)\ \nonumber \\
&=&2M^2(\gamma^{\mu}\partial_{\mu}S(x)
-\nabla^{\mu}V_{\mu}(x) -\gamma_{5}\Sigma^{\mu\nu}\nabla_{\mu}A_{\nu}(x))
+M\Sigma^{\mu\nu\lambda}\nabla_{\mu}F_{\nu\lambda}(x) , \nn \\ 
&=&0 , \label{sum}
\end{eqnarray}
where the third rand anti-symmetric tensor of gamma matrices is denoted: 
$ 
\Sigma^{\mu\nu\lambda}=(\gamma^{\mu\nu\lambda}+\gamma^{\lambda\mu\nu}+\gamma^{\nu\lambda\mu}
-\gamma^{\mu\lambda\nu}-\gamma^{\nu\mu\lambda}-\gamma^{\lambda\nu\mu})/3!
$. 
Here we should check the consistency between he independent terms in the sum BW equations (\ref{sum})  
and the subtraction BW equations (\ref{Peq})-(\ref{Feq}).  
Using the subtraction BW equations we can derive the following relations 
\begin{eqnarray}
\gamma^{\mu}\partial_{\mu}S(x)&=&0 , \label{sumS} \\ 
M^2 \nabla^{\mu}V_{\mu}(x)&=&[\nabla^{\nu} , \nabla^{\mu}] \nabla_{\mu}V_{\nu}(X) 
=C^{\lambda}_{\ , \mu\nu} \nabla_{\lambda}\nabla^{\mu} V^{\nu}(x ), \label{sumV}  
\\ 
\Sigma^{\mu\,\nu}\nabla_{\mu}A_{\nu}(x)
&=&\frac{1}{2}\Sigma^{\mu\nu}C^{\lambda}_{\ ,\mu\nu}\partial_{\lambda}P(x)\ , \label{sumA}   \\  
\Sigma^{\mu\nu\lambda}\nabla_{\mu}F_{\nu\lambda}(x)
&=&\Sigma^{\mu\nu\lambda}(-R^{\rho}_{\lambda\ , \mu\nu}V_{\rho}(x) 
+C^{\rho}_{\ ,\mu\nu}\nabla_{\rho}V_{\lambda}(x)) \ , \label{sumF}
\end{eqnarray} 
where $C^{\rho}\ , _{\mu\nu}=\Gamma^{\rho} _{\mu\nu}-\Gamma^{\rho} _{\nu\mu}$ 
and  $R^{\rho}_{\lambda\ , \mu\nu}$ denote the torsion tensor  and  the Riemann curvature tensor. 
The proof of the consistency equations (\ref{sumS})-(\ref{sumF}) is given in Appendix A. 
The lower suffix in the curvature tensor are totally antisymmetric because of the factor $\Sigma^{\mu\nu\lambda} $
and the identity relation holds  
\begin{eqnarray}
R^{\rho}_{[\lambda\ , \mu\nu]}
=-\nabla_{[\mu}C^{\rho}_{\ ,\nu\lambda]}-C^{\tau}\ , _{[\mu\nu}C^{\rho}\ , _{\lambda]\tau} , 
\end{eqnarray}
where  the anti-symmetrized suffix notation in square bracket is defined:  
\begin{eqnarray}
A_{[\mu\nu\lambda]}:=\frac{1}{3!}(A_{\mu\nu\lambda}+A_{\lambda\mu\nu}
+A_{\nu\lambda\mu}-A_{\mu\lambda\nu}-A_{\nu\mu\lambda}-A_{\lambda\nu\mu}) . \label{asymmetrized}
\end{eqnarray}
In order to satisfy the sum of the BW equations, 
we require the torsion-free condition for the background spacetime geometry:
\bea 
C^{\rho}\ ,_{\mu\nu}=\Gamma^{\rho}_{\mu\nu}-\Gamma^{\rho}_{\nu\mu}=0 \label{torsionfree} . 
\eea

%%%%%%%%%%%%%%%%%%%%   bi-spinor expressed by P and V  %%%%%%%%%%%%%%%%%

Inserting the independent Boson fields of pseudoscalar and vector fields in the bi-spinor field expression 
in eq. (\ref{biB}), we find 
\bea
\Psi(x)=M(M-\gamma^{\mu}\nabla_{\mu})(\gamma_{5}P(x)+\gamma^{\nu}V_{\nu}(x)) , \label{BWsol}
\eea 
 with the supplementary condition, which is from eq. (\ref{sumV}) with the torsion free condition (\ref{torsionfree}), 
\bea  \nabla^{\nu}V_{\nu}(x)=0 . \label{supplementary}\eea 
The equation eq. (\ref{BWsol}) shows the direct relation between bi-spinor and Boson fields. 

We note that the transpose operation with the charge conjugation $C$ for  pseudoscalar and vector fields 
show even and odd properties respectively: 
\bea
C[(M-\gamma^{\mu}\nabla_{\mu})\gamma_{5}P(x)]^{T}C^{-1}&=&(M-\gamma^{\mu}\nabla_{\mu})\gamma_{5}P(x) , \nn \\
C[(M-\gamma^{\mu}\nabla_{\mu})\gamma^{\nu}V_{\nu}(x)]^{T}C^{-1}&=&
-(M-\gamma^{\mu}\nabla_{\mu})\gamma^{\nu}V_{\nu}(x) . 
\eea
This even and odd  property corresponds to their spins 0 and 1 for pseudoscalar and vector fields respectively.

%%%%%%%%%%%%%%%%%%%%%%%%%%%%%%%%%%%%%%%%%%%%%%%%%%%%%%%%%%%%%%%%%%%%
%%%%%%%%%%%%%%%%%%%%%   F-B correspondence %%%%%%%%%%%%%%%%%%%%%%%%%%%
\section{Fermion-Boson correspondence and spin structure}

Next we will show to express bi-spinor fields as a direct product of two spinor fields. 
Then we will derive the direct relations between spinors and Boson fields.  

%%%%%%%%%%  Dirac equations in local inertial coordinate system %%%%%%%%%%%%%

The general curved coordinate system is related to the local inertial coordinate system by the 
vierbeins $e_{a}^{\mu}$.  
The BW equations for bi-spinors in the local inertial coordinate system 
are the Dirac type equations in eqs. (\ref{BWinM}) as   
\[
(\gamma^{a}\partial_{a}+M)\Psi(x)=0  , \ \ \mbox{and} \ \   
\Psi(x) (\overleftarrow{\partial}_{a}\gamma^{a}-M)=0  . 
\]
Firstly we consider the pseudoscalar case. 
The bi-spinor field can be obtained from the BW solution in eq. (\ref{BWsol}) for the pseudoscalar field as
\bea
\Psi_{P}(x)=M(M-\gamma^{a}\partial_{a})\gamma_{5}P(x) . 
\eea
An analytic pseudoscalar field $P(x)$ can be expanded in the Fourier series with momentum $k^a$: 
\bea
P(x)=\sum_{k}P(k) \ \ \ \mbox{with} \ \ \ P(k)\equiv c_{k}\exp(ik^ax_a) , \label{P(k)}
\eea
where $c_{k}$ is the expansion coefficients. 
For each plane wave $P(k)$, the bi-spinor field can be decomposed into 
sets of ket spinors and bra spinors as
\bea
|P(k)^j>&\equiv &(M/2-\gamma^a\partial_a)\sqrt{2P(k)}|j> , \label{defket} \\
<P(k)^j|&\equiv &<j|\sqrt{2P(k)}(M/2-\overleftarrow{\partial_a}\gamma^a)\gamma_{5} , \label{defbra}
\eea
where the ket and bra unit spinors are the base of a complete set: 
\bea
|1>=\left( \begin{array}{c} 1 \\ 0 \\ 0 \\ 0 \end{array} \right),  \ 
|2>=\left( \begin{array}{c} 0 \\ 1 \\ 0 \\ 0 \end{array} \right),  \ 
|3>=\left( \begin{array}{c} 0 \\ 0 \\ 1 \\ 0 \end{array} \right),  \
|4>=\left( \begin{array}{c} 0 \\ 0 \\ 0 \\ 1 \end{array} \right) , \ \label{unit}
\eea
with $\Sigma_{j=1}^{4}|j><j|=1$. 
The ket and bra spinors defined in eqs.(\ref{defket})-(\ref{defbra}) satisfy the Dirac equation of mass $M/2$: 
\bea
(\gamma^a\partial_a +M/2)|P(k)^j>=0 \ \ \mbox{and} \ \ <P(k)^j|(\overleftarrow{\partial_a}\gamma^a-M/2)=0 .
\eea
Then we can read the energy  and spin  dependence for each ket spinor from the expression in eq.(\ref{defket}) as
 \bea
|P(k)^1>&=&(M+E)|u_ {\uparrow}(k^{a})>\sqrt{P(k)/2} , \nn \\ 
|P(k)^2>&=&(M+E)|u_{\downarrow}(k^{a} )>\sqrt{P(k)/2} , \nn \\ 
|P(k)^3>&=&(M-E)|v_{\downarrow}(-k^{a})>\sqrt{P(k)/2} , \nn \\ 
|P(k)^4>&=&-(M-E)|v_{\uparrow}(-k^{a})>\sqrt{P(k)/2} , \label{ket} \eea
where spinors or anti-spinors are denoted by $u$ and $v$, spins up or down  
are denoted by  $\uparrow$ or $\downarrow$ respectively, and four momentum is denoted as $k^{a}=(E, k_{1}, k_{2}, k_{3})$. 
The charge conjugation relates them as 
\bea
C: \gamma_2(|u_{\uparrow}(k^a)>)^*=|v_{\uparrow}(k^a)> \ \ ,  \ \ 
\gamma_2(|u_{\downarrow}(k^a)>)^*=|v_{\downarrow}(k^a)> . \label{charge}
\eea
Similarly for each bra spinor we obtain from the definition in eq. (\ref{defbra}) as
\bea
<P(k)^1|&=&-(M+E)<v_{\downarrow}(\tilde{k}^{a})|\sqrt{P(k)/2} ,\nn  \\
<P(k)^2|&=& (M+E)<v_{\uparrow}(\tilde{k}^{a})|\sqrt{P(k)/2} , \nn \\
<P(k)^3|&=&-(M-E)<u_{\uparrow}(-\tilde{k}^{a})|\sqrt{P(k)/2} , \nn \\
<P(k)^4|&=&-(M-E)<u_{\downarrow}(-\tilde{k}^{a})|\sqrt{P(k)/2} . \label{bra}
\eea
The ket and bra spinors are related by the transpose operation as 
\bea
(|u_{\uparrow}(k^a)>)^T=<u_{\uparrow}(k^a)| \ \ ,  \ \ 
(|u_{\downarrow}(k^a)>)^ T=<u_{\downarrow}(k^a)| ,
\eea
and similar to the anti-spinors.
The modified four momentum in eqs.(\ref{bra}) is denoted as $\tilde{k}^{a}=(E, -k_{1}, k_{2}, -k_{3})$. 
where the sign of momenta of $k_{1}$ and $k_{2}$ is changed due to the transpose property of 
the gamma matrices: $\gamma^{1\ T}=-\gamma^{1}$ and  $\gamma^{2\ T}=-\gamma^{2}$ 
to express bra spinors from ket spinors. 
The explicit expression for the ket and bra spinors for psuedoscalar fields is in Appendix B. 

Using these spinors, the bi-spinor for the pseudoscalar field is recovered: 
\bea
&&\sum_{j=1}^{4}|P(k)^j><P(k)^j| \nn \\
&=&\{(M+E)^2(|u_{\uparrow}(k^{a})><v_{\downarrow}(\tilde{k}^{a})
| -|u_{\downarrow}(k^{a})><v_{\uparrow}(\tilde{k}^{a})| ) \nn \\
&&+(M-E)^2(|v_{\downarrow}(-k^{a})><u_{\uparrow}(-\tilde{k}^{a})| 
  -|v_{\uparrow}(-k^{a})><u_{\downarrow}(-\tilde{k}^{a})| ) \}P(k)/2 , \nn \\
&=&\ \ \ M(M-\gamma^a\partial_a )\gamma_{5}P(k) ,  \label{decomposition}
\eea
where we have used the field equation for the psuedoscalar field (\ref{Peq}) and  the 
identity relation: 
\bea
(\gamma^a\partial_a+M)^2\gamma_5P(x)=2M(\gamma^a\partial_a+M)\gamma_5P(x) .
\eea
From the spinor decomposition equation (\ref{decomposition}), we can recognize 
that direct product of spinors and anti-spinors forms a spin 0 pseudoscalar field. 

In the parallel way, we can confirm that the direct product of spinors and anti-spinors 
forms a spin 1 vector field, whose proof is in Appendix C. 

Therefore the BW formulation is a kind of dynamical realization of algebraic direct product representation 
of rotation group: 
\bea
1/2\otimes1/2^- =0^-\oplus 1^- , \label{rotationgroup}
\eea
where the numbers and the minus sign show the magnitude of spins and their parity.   

Based on the BW formulation, one of the important statements on the Fermion-Boson correspondence 
is the superradiant problem in the black hole background geometry. 
If the Bose particles behave like superradiant instability, the Fermi particles should behave 
like superradiant instability too, and vice versa under the Fermion-Boson correspondence.     
 
In the next section, we shall study to make clear the superradiant problem for both of Fermi and Bose particles 
in the view of particle spectra.

%%%%%%%%%%%%%%%%%%%% spectrum condition and superradiant phenomena %%%%%%%%%%%%%%%%%%%%%%%%%%%%

\section{Spectrum condition in co-rotating coordinate system and superradiant problem}
\subsection{Spectrum condition in co-rotating coordinate system}
\setcounter{equation}{0}

In this subsection, we try to derive the spectrum condition for Fermi and Bose particles around rotating geometry 
as radiant phenomena.  The (3+1)-dimensional spacetime is assumed to be stationary and axially symmetric 
for which the metric tensors are independent of the time $t$ and the azimuthal angle $\phi$.  
The invariant length is expressed in the polar coordinate as  
\bea
ds^2=g_{tt}(r,\theta)dt^2+2g_{t\phi}(r,\theta)dtd\phi+g_{\phi\phi}(r,\theta)d\phi^2
+g_{rr}(r,\theta)dr^2+g_{\theta\theta}(r,\theta)d\theta^2. 
\eea  
In order to study the co-rotating coordinate system, 
which is locally non-rotating observer system \cite{Bardeen1970,MTW1970},    
we make to diagonalize the $t-\phi$ part of the invariant length: 
\bea
ds_{t-\phi}^2= g_{tt}dt^2+g_{t\phi}dtd\phi +g_{\phi\phi}d\phi^2
\equiv g_{\bar{t}\bar{t}}d\bar{t}^{2}+g_{\bar{\phi}\bar{\phi}}d\bar{\phi}^2,
\eea  
where $\bar{t}$ and $\bar{\phi}$ denote the time and azimuthal angle in co-rotating coordinate system respectively.  
The local co-rotation coordinate system and the general rotation spacetime are related by the linear transformation: 
\bea
(d\bar{t}, d\bar{\phi})=(dt, d\phi)U,
\eea
and the metric tensors are related as 
\bea
\bar{G}=\left(\begin{array}{cc} g_{\bar{t}\bar{t}}  & 0 \\ 0 & g_{\bar{\phi}\bar{\phi}} \end{array}\right) 
=U^{-1}G(U^{T})^{-1} \ , \ 
G=\left(\begin{array}{cc} g_{tt} & g_{t\phi} \\ g_{t\phi} & g_{\phi\phi} \end{array}\right) \label{Grelation}
\eea
with the transfer matrix:  
\bea
U=\left( \begin{array}{cc}
1 & X(r,\theta) \\
Y(r,\theta) & 1 
\end{array}\right), 
\eea
where the diagonal element is set to one as our normalization 
and elements $X$, $Y$ are to be determined taking account of their mass dimensions dim[$X$]=1, dim[$Y$]=-1.  

As to the expressions for the transfer matrix elements, 
$X$ is chosen to obtain the local co-rotating coordinate system 
adjusting the angular velocity of rotating spacetime \cite{Bardeen1970,MTW1970}: 
\bea
X=-\Omega(r,\theta)= {g_{t\phi}}(r,\theta)/ {g_{\phi\phi}}(r,\theta), \label{co-rot}
\eea
where  $\Omega(r,\theta)\, (:=d\phi/dt)$ denotes the angular velocity of spacetime.  
From and the $\bar{G}$ and $G$ relation of eq.(\ref{Grelation}) and the co-rotating condition of eq.(\ref{co-rot}), 
we obtain another matrix element as 
\bea
Y=0 ,
\eea 
and the $t-\phi$ part of the invariant length is  
\bea
ds_{t-\phi}^2=(g_{tt}-\frac{g_{t\phi}^2}{g_{\phi\phi}})dt^2+g_{\phi\phi}(d\phi+\frac{g_{t\phi}}{g_{\phi\phi}}dt)^2
\eea 

In addition to the coordinate forms, the differential operators are introduced to form an invariant quantity:    
\bea
dt{\partial}/{\partial_t} + d\phi{\partial}/{\partial_{\phi} }
=d\bar{t}{\partial}/{\partial_{\bar{t}}} + d\bar{\phi}{\partial}/{\partial_{\bar{\phi}} },
\label{invariant}\eea
where the differential operators transform as  
\bea
({\partial}/{\partial_{\bar{t}}}, {\partial}/{\partial_{\bar{\phi}} })=
({\partial}/{\partial_t}, {\partial}/{\partial_{\phi} })U.
\eea
Because metric tensors are independent of the time $t$ and the azimuthal angle $\phi$, 
the $t-\phi$ factor of matter fields is the exponential form as  
\bea
\Phi\propto \exp{(-i\omega t+im\phi)}, \label{matter}
\eea
where the matter field $\Phi$ stands for bi-spinor fields $\Psi$, pseudoscalar fields $P$, vector fields $V_{\mu}$,  
spinors, and anti-spinors explained in the previous sections.  
The notations $\omega$ and $m$ denote the frequency as a field (the energy as a particle) 
and azimuthal angular momentum respectively. 
Applying the differential operator to matter fields of the form eq.(\ref{matter}), the invariant quantity of eq.(\ref{invariant}) becomes
\bea 
-\omega dt+md\phi= -\bar{\omega}d\bar{ t} + \bar{m}d\bar{ \phi},  
\eea
where 
\bea
\bar{\omega}=\omega+Xm\ , \ \bar{m}=m.
\eea
As to the stationary and axially symmetric spacetime, the reflection symmetry on $t-\phi$ part holds:   
$(dt, d\phi)\rightarrow -(dt, d\phi)$. 
And by this symmetry, the reflection symmetry on the spectrum also holes: 
\bea
({\omega}, {m}) \leftrightarrow -({\omega}, {m}) , \ \ \ 
(\bar{\omega}, \bar{m}) \leftrightarrow -(\bar{\omega}, \bar{m}) \label{reflection}
\eea 
in general rotating geometry and in co-rotating coordinate system respectively.  
On the frequency, one side of the spectra is physical modes and other is unphysical modes. 
The physical normal modes are half of the full spectrum and the physical modes 
in co-rotating coordinate system are chosen as    
\bea
\bar{\omega}>0, 
\eea 
which naturally connect the non-rotating coordinate systems. 
We call the spectrum condition for the radiant processes in the rotating spacetime.  

For the explicit expression for the metric tensors, we take the Kerr metric, which is 
 \bea
g_{tt}&=&\frac{1}{\rho^2}(-\Delta+a^2\sin^2{\theta}), \ \ \
 g_{t\phi}=\frac{a\sin^2{\theta}}{\rho^2}(\Delta-(r^2+a^2))
\nn \\
g_{\phi\phi}&=&\frac{\sin^2\theta}{\rho^2}(-a^2\sin^2\theta\Delta+(r^2+a^2)^2), 
\ \ \ g_{rr}=\frac{\rho^2}{\Delta}, \ \ \ g_{\theta\theta}={\rho^2}
\eea
with  
\bea
\Delta&=&r^2+a^2-2M_{BH}r , \ \ \  \rho=r^2+a^2\cos^2{\theta},
\eea
where $M_{BH}$ and $a$ denote the mass and angular momentum per unit mass of the black hole.    
%%%%%%%%%%%%%%%%%%%%%%%%%%%%%%% footnote 1 %%%%%%%%%%%%%%%%%%%%%%%%%%%%%%%%%%%%%%%%%%%%%%%%%%%%%
\footnote{We note that 
the geodesic motion of a particle shows the Coriolis force in an asymptotic region: 
\bea
d\bar{\phi} \rightarrow d\phi -{2aM_{BH}dt}/{r^3}, \ \ \ (a, 2M_{BH} \ll r), 
\eea 
which support our choice of co-rotating coordinate system of eq.(\ref{co-rot}).
}
%%%%%%%%%%%%%%%%%%%%%%%%%%%%%%%%%%%%%%%%%%%%%%%%%%%%%%%%%%%%%%%%%%%%%%%%%%%%%%%%%%%%%%%%%%%%%%
Near the outer horizon $r=r_{H}:=M_{BH}+({M_{BH}^2-a^2})^{1/2}$, the spectrum condition becomes 
\bea \omega-m \Omega_{H} >0, \label{SConH}
\eea 
where $\Omega_{H} =a/(r_{H}^2+a^2)$ denotes the angular velocity of black hole near the horizon. 
%%%%%%%%%%%%%%%%%%%%%%%%% footnote 2 %%%%%%%%%%%%%%%%%%%%%%%%%%%%%%%%%%%%%%%%%%%%%%%%%%%%%%%%%%%%%%
\footnote{
It is noted that another interesting choice of co-rotating coordinate system is to put the angular velocity with 
the horizon condition ($\Delta=0$) to the matrix element in the Kerr metric:  
\bea
X'=-\Omega'=g_{t\phi}/g_{\phi\phi}\mid_{\Delta=0}=-a/(r^2+a^2),
\eea
where the new choice of co-rotating coordinate system is denoted by the prime suffix. 
In this choice, new coordinate system is obtained as  
\bea
Y'&=&(X'g_{\phi\phi}-g_{t\phi})/(X'g_{\phi t}-g_{tt})=-a\sin^2{\theta}, \nn \\
dt'&=&dt-{a\sin^2{\theta}}d\phi, \ \ \ d\phi'=d\phi-{adt}{/(r^2+a^2)},
\eea
and matrix tensors are 
\bea
g_{tt}'=-\Delta/\rho^2, \ \ \ g_{\phi\phi}'=\sin^2{\theta}(r^2+a^2)^2/\rho^2. 
\eea
This coordinate system becomes the Boyer-Lindquist coordinates \cite{BL1967}:
\bea
ds_{t-\phi}^2&=&g_{tt}'dt'^2+g_{\phi\phi}'d\phi'^2\nn\\
&=&-{\Delta}(dt-a\sin^2{\theta}d\phi)^2/{\rho^2}
+{\sin^2{\theta}}((r^2+a^2)d\phi-adt)^2/\rho^2.
\eea
The metric condition on the horizon is the same form as in eq.(\ref{SConH}) in this co-rotating coordinate system. 
}
%%%%%%%%%%%%%%%%%%%%%%%%%%%%%%%%%%%% Tortoise coordinate %%%%%%%%%%%%%%%%%%%%%%%%%%%%%%%%%%%%
As another understanding of the spectrum condition (\ref{SConH}), we can express the radial part of the wave function as
\bea
\Phi_{radial}\simeq \exp{ [- i(\omega-m\Omega_H)r_*]  },
\eea 
where the radial tortoise coordinate  \cite{RW1957} is defined $dr_*=(r^2+a^2)\Delta^{-1}dr$ 
and  the radial momentum is denoted by $\omega-m\Omega_H$ 
as the positive value is for in-going wave to the black hole.   
Therefore we analyze the superradiant problem under the consideration of the spectrum condition 
relating the role in the frequency $\omega$ and the radial momentum $\omega-m\Omega_H$ 
in the next subsection.  

%%%%%%%%%%%%%%%%%%%%%%%%%%%%%%%%%%%%%%% superradiant problem %%%%%%%%%%%%%%%%%%%%%%%
\subsection{Superradiant problem}

The superradiant problem is one of the longstanding problems and is still important nowadays. 
In this subsection, we study the superradiant problem taking account the Fermion-Boson correspondence 
and the spectrum condition. 
 
%%%%%%%%%%%%%%%%%%%%%%%%% Bose SR %%%%%%%%%%%%%%%%%%%%%%%%%%%%%%%%%%%%%%%%%%%%%%%%%% 
For the Bose particles, the superradiant condition was derived from the current conservation 
near the rotating black hole horizon \cite{Unruh1974,Press1972,Misner1972,Detweiler1980,Takasugi1997, Mukohyama2000}:  
\bea
\omega-m\Omega_{H}<0  \ \ {\mbox with} \ \ \ \omega>0, \label{Type1}
\eea  
which we call the type 1 superradiance.  
Especially for the small rotating black hole and the positive incident particle energy, 
the superradiant instability are studied to occur under the condition of (\ref{Type1})
\cite{Press1972, Chandrasekhar1983,Cardoso2004,Kodama2008}. 
As we pointed out in the previous subsection in eq.(\ref{reflection}), 
the particles have the reflection symmetry on the spectrum 
$(\omega, m) \rightarrow -(\omega, m)$, there is another superradiant possibility:   
\bea
\omega-m\Omega_{H}>0 \ \ \  {\mbox with } \ \ \ \omega<0,  \label{Type2}
\eea 
which we call the type 2 superradiance. 
The type 2 superradiant condition is consistent with our spectrum condition in eq.(\ref{SConH}).    

%%%%%%%%%%%%%%%%%%%%%%%%%%% Fermi SR %%%%%%%%%%%%%%%%%%%%%%%%%%%%%%%%%%%%%%%%
On the other hand, for Fermi particles, the type 1 superradiance has been studied not to occur
\cite{Unruh1974, Maeda1976, Wagh1985}. 
One of the reasons is the spectrum region of type 1 is occupied by the Fermion particles and 
cannot realize in this spectrum region because of the Fermi statistics. 
We note that there is another possible superradiant spectrum region of type 2, 
because this spectrum region is not occupied and possible to occur.     

We also note that the type 2 superradiance is possible for both Bose and Fermi particles 
and the result is consistent our previous study on Fermion-Boson correspondence and the spectrum condition. 
The type 2 superradiance is the necessary and stable spectrum modes 
because of the reflection symmetry with respect to $\omega$ and $m$.   
As black holes would not recognize particles as elementary or composite, 
the similar radiative behavior for the Fermi and Bose particles to black holes is desirable 
based on the Fermion-Boson correspondence.    
  
%%%%%%%%%%%%%%%%%%%%%%%%%%%%%%%%%%%%%%%%%%%%%%%
%%%%%%%%%%%%%%%%%%%%%%%%%%%%%%%%%%%%%%%%%%%%%%%
\section{Summary and discussion}
\setcounter{equation}{0}

We have studied the interacting relations of matter fields with the curved spacetime. 
We obtained the Bargmann-Wigner formulation on general curved spacetime, 
which describe the bi-spinors as pseudoscalar and vector fields consistently as in eq.(\ref{BWsol}) under 
the torsion-free condition in eq.(\ref{torsionfree}) for the background spacetime geometry.   
We also derived the direct relations between Fermi (spinor and anti-spinors) 
and Boson (pseudoscalar and vector) fields via bi-spinor fields. 
The spin structures show that the BW formulation is a kind of dynamical realization 
of the representation under rotation group as quark and anti-quark system forms a 
$\pi$- and $\rho$- mesons as in eq.(\ref{rotationgroup}).
 Following this Fermion-Boson correspondence, 
we investigated the field spectrum around the rotation black hole  and obtained the spectrum condition 
$\omega-m\Omega_H>0$ in eq.(\ref{SConH})
for both Boson and Fermion fields in the co-rotating coordinate system as locally non-rotating observer
 (\ref{co-rot}).  
We also applied these results to the superradiant problem in the Kerr spacetime geometry 
and obtained the result that the type 2 superradiance is possible to occur as
$\omega-m\Omega_H>0$ with $\omega<0$ in  eq.(\ref{Type2}). 

These results are consistent with our previous research works based on the BW formulation  
 \cite{KC2015} and the quantum field theory \cite{KCSY2017} in (3+1)-dimensional spacetime 
and the BTZ spacetime in (2+1)-dimension \cite{Kenmoku2008,Kenmoku2008-2}.  

%%%%%%%%%%%%%% Bogoluibov transformation %%%%%%%%%%%%%%%%%%
The type 2 superradiance is the positive radial momentum $(\omega-m\Omega_H>0)$ with 
negative energy $(\omega<0)$ modes. 
We can interpret this phenomenon of 
the annihilation of anti-particle with negative energy into a black hole  
as the creation of particle with positive energy from the black hole. 
This process can be expressed by the Bogoliubov transformation for Fermi particles: 
\bea
U_{\theta}=\exp{[\theta \sum_{\alpha+\beta=0} (a_{\alpha}b_{\beta}-b_{\beta}^{\dagger}a_{\alpha}^{\dagger})]},
\eea
where $a_{\alpha}, b_{\beta}$ denote the particle, anti-particle annihilation operators with 
positive or negative quantum number $\alpha=-\beta=(\omega, m, \dots) $ respectively 
under the spectrum condition (\ref{SConH}).  
The mixing parameter $\theta$ is proportional to the angular velocity of the black hole $\Omega_H$. 
The dressed anti-particle operator is obtained by the pair creation effect as
$ 
\hat{b}_{\beta}:= U_{\theta} b_{\beta}U_{\theta}^{\dagger}
=\cos{\theta}\, b_{\beta}+\sin{\theta}\, a_{\alpha}^{\dagger}. 
$ 
The annihilation of the anti-particle into black hole shows that 
the creation of particle with positive energy from the black hole as 
\bea
H\hat{b}_{\beta}|0>=\sin{\theta}\, \omega\, a_{\alpha}^{\dagger}|0>, \ \ (\alpha+\beta=0), \label{sintheta}
\eea
where the Hamiltonian operator is 
$H=\sum \omega (a_{\alpha}^{\dagger}a_{\alpha}-{b^{\dagger}}_{\beta}b_{\beta})$ 
and the vacuum represents the black hole state as $b_{\beta}|0>=0$. 
Similarly the annihilation of particle with negative energy into a black hole 
can be interpreted as the creation of anti-particle with positive energy from the black hole. 

For Boson particles, the mixing angle should be replaced: 
$\sin{\theta} \rightarrow {\mbox{sinh}}\, {\theta}$ in eq.(\ref{sintheta}).  
By this mechanism, the energy can be transferred to Fermi and/or Bose particles 
from rotating black holes.    
This is known as the Penrose mechanism to extract the black hole energy in the ergo region
\cite{Penrose1969}.  

Combination the type 2 superradiant modes ($\omega-m\Omega_H>0$ with $\omega<0$) 
and standard normal modes ($\omega-m\Omega_H>0$ with $\omega>0$) 
forms a complete set of physical normal modes both for Fermion and Boson 
independent of elementary or composite particles \cite{KCSY2017}. 
%%%%%%%%%%%%%%%%%%%%%%%%%%%%%%%%%%%%%%%%%%%%%%%%%%%%%%%%
\section*{Acknowledgements}

Author would like to give thanks to Professor K. Shigemoto 
 for his encouragement and discussions to this manuscript. 

%%%%%%%%%%%%%%%%%%%%%%%%%%%%%%%%%%%%%%%%%%%%%%%%%%%%%%%%
\section*{Appendix}

\appendix
%%%%%%%%%%%%%%%%%%%%%% Appendix A %%%%%%%%%%%%%%%%%%%%%%%%%%%%%

\section{ Proof of the consistency equations (\ref{sumS})-(\ref{sumF}) }
\setcounter{equation}{0}
We shall show the consistency equations step by step in the following. 
\begin{itemize}
\item[1] The proof of equation  (\ref{sumS})\\
The equation  $\gamma^{\mu}\partial_{\mu}S(X)=0$ is simply by the equation  $S(x)=0$ in eq. (\ref{Seq}): .  
\item[2] The proof of equation (\ref{sumV}) \\ 
For the general tensor field $T^{\rho\sigma}(x)$, we can show the relation: 
\bea
[\nabla_{\nu} , \nabla_{\mu} ] T^{\rho\sigma}(x)=-R^{\rho}_{\ \lambda , \mu\nu}T^{\lambda\sigma}(x) 
-R^{\sigma}_{\ \lambda , \mu\nu}T^{\lambda\rho}(x)+C^{\lambda}_{\ , \nu\mu}\nabla_{\lambda}T^{\rho\sigma}(x) , 
\eea 
where $C^{\lambda}_{\ , \nu\mu}$ is the torsion tensor. 
Setting the index parameters as $\rho=\mu$ and $\sigma=\nu$, we obtain 
\bea
[\nabla_{\nu} , \nabla_{\mu} ] T^{\mu\nu}(x)=C^{\lambda}_{\ , \nu\mu}\nabla_{\lambda}T^{\mu\nu}(x) .
\eea 
Identifying $T^{\mu\nu}=\nabla^{\mu}V_{\nu}(x)$ in this formula   
\bea
M^2 \nabla^{\mu}V_{\mu}(x)=C^{\lambda}_{\ , \mu\nu} \nabla_{\lambda}\nabla^{\mu} V^{\nu}(x),
\eea
the equation (\ref{sumV}) is proved. 
\item[3] The proof of equation (\ref{sumA}) \\
From the expression for the axial vector $A_{\mu}(x)$ in eq. (\ref{Aeq}), we obtain 
\bea
2\Sigma^{\mu\nu}\nabla_{\mu}A_{\nu}(x)&=&[\nabla_{\mu} , \nabla_{\nu}]P(x) \nn \\
			&=&\Sigma^{\mu\nu}(\Gamma^{\lambda}_{\ \mu\nu
}-\Gamma^{\lambda}_{\ \nu\mu} )\partial_{\lambda}P(x) .
\eea
Using the relation $\Gamma^{\lambda}_{\ \mu\nu}-\Gamma^{\lambda}_{\ \nu\mu}
=C^{\lambda}_{\ \mu\nu} $, we have proved the equation (\ref{sumA}). 
\item[4] The proof of the equation (\ref{sumF})\\
From the expression for tensor field in eq. (\ref{Feq}),  we obtain 
\bea
\Sigma^{\mu\lambda\tau}\nabla_{\mu}F_{\lambda\tau}(x)
=\Sigma^{\mu\lambda\tau} [\nabla_{\mu}, \nabla_{\lambda}] V_{\tau}(x) .
\eea
Using the formula $
 [\nabla_{\mu}, \nabla_{\lambda}] V_{\tau}(x)=-R^{\rho}_{\ \tau , \mu\lambda}V_{\rho}(x)
+C^{\lambda}_{\ ,\mu\nu}\nabla_{\lambda}V_{\sigma}(x)
$, 
we get 
\bea
\Sigma^{\mu\lambda\tau}\nabla_{\mu}F_{\lambda\tau}(x)
=\Sigma^{\mu\lambda\tau} (-R^{\rho}_{\ \tau , \mu\lambda}V_{\rho}(x)
+C^{\lambda}_{\ ,\mu\nu}\nabla_{\lambda}V_{\sigma}(x)) .
\eea
Because of the factor $\Sigma^{\mu\lambda\tau}$  
the lower suffix $\tau, \mu, \lambda$ in $R^{\rho}_{\ \tau , \mu\lambda}$ can be totally antisymmetrized, 
and the equation (\ref{sumF}) has been proved.  

\end{itemize}    

%%%%%%%%%%%%%%%%%%%%%%% Appendix B %%%%%%%%%%%%%%%%%%%%%%%%%%%%%%%%%%%%%%%%
\section{The ket and bra spinors for psuedoscalar fields}
\setcounter{equation}{0}

Gamma matrices in local inertial coordinate system (flat Minkowski spacetime) is represented: 
\bea
  &&\gamma^{0}=-i \left(\begin{array}{cc}  
  I & 0\\
  0 & -I
  \end{array} \right),  \quad
  \gamma^{a}=\left(\begin{array}{cc}  
  0 & -i\vec{\sigma}\\
  i\vec{\sigma} & 0
  \end{array} \right),  
\eea
where $\vec{\sigma} $ stands for Pauli matrices. 

For ket spinors defined in eq. (\ref{defket}), the explicit matrix expression is obtained as 
\bea
(M/2-\gamma^a\partial_{a})\sqrt{2P(k)}|j>
&=&(M-i\gamma^a k_a)|j>\sqrt{P(k)/2} , \nn \\
&=&\left(
\begin{array}{cc}
M+E & -\vec{\sigma} \cdot \vec{k} \\
\vec{\sigma} \cdot \vec{k} & M-E 
\end{array}
\right)|j>\sqrt{P(k)/2} ,
\eea
where $P(k)=c_{k}\exp{(ik^ax_a)}$ in eq.(\ref{P(k)}). 
From this expression we can read each component of the ket spinors and anti-spinors in eqs. (\ref{ket}) as 
\bea
|u_{\uparrow}(k^a)>&=&
\left( \begin{array}{c}
1 \\ 0 \\ k_{3}/(M+E) \\ (k_{1}+k_{2})/(M+E) 
\end{array} \right) , 
|u_{\downarrow}(k^a)>=
\left( \begin{array}{c}
0 \\ 1 \\ (k_{1}-ik_{2})/(M+E) \\ -k_{3}/(M+E) 
\end{array} \right) , \nn \\
|v_{\downarrow}(-k^a)>&=&
\left( \begin{array}{c}
- k_{3}/(M-E) \\  -(k_{1}+k_{2})/(M-E) \\ 1 \\ 0
\end{array} \right) , 
|u_{\uparrow}(-k^a)>=-
\left( \begin{array}{c}
 -(k_{1}-ik_{2})/(M-E) \\ k_{3}/(M-E) \\ 0 \\1 
\end{array} \right) ,
\eea 
where spinors $u$ and anti-spinors $v$ are charge conjugation pair each other defined in eqs.(\ref{charge}).   

For bra spinors defined in eq. (\ref{defbra}), the explicit matrix expression is obtained as 
\bea
%<P(k)^j|&\equiv&
\sqrt{2P(k)}<j|(M/2-\overleftarrow{\partial_a}\gamma^a)\gamma_{5} 
&=&<j|(M-i\gamma^a k_a)\gamma_5\sqrt{P(k)/2} , \nn \\
&=&-<j|\left(
\begin{array}{cc}
 -\vec{\sigma} \cdot \vec{k}  & M+E\\
M-E & \vec{\sigma} \cdot \vec{k}  
\end{array}
\right)\sqrt{P(k)/2} . 
\eea
Here we know that the ket and bra relation is 
$u_{\uparrow}(k^a)\rightarrow -v_{\downarrow}(\tilde{k}^a)$ 
and $u_{\downarrow}(k^a)\rightarrow v_{\uparrow}(\tilde{k}^a)$  by the effect of the operation $\gamma_{5}$. 
The modified four momentum $\tilde{k}^a$ is explained below eq. (\ref{bra}).  
Then we obtain the expression for bra spinors in eq. (\ref{bra}).  

%%%%%%%%%%%%%%%%%%%%%%% Appendix C %%%%%%%%%%%%%%%%%%%%%%%%%%%%%%%%%%%%%%%%
\section{The ket and bra spinors for vector fields}
\setcounter{equation}{0}

For the vector fields in the curved spacetime, 
we can set up a local inertial coordinate system similar to the pseudoscalar case. 
The BW solution for the vector fields is from eq. (\ref{BWsol}) as 
\bea
\Psi_V(x)=M(M-\gamma^a\partial_a)\gamma^aV_a(x) .
\eea
The analytic solution for the vector fields can be expanded in Fourier series, 
\bea
V_a(x)=\sum_{k, I} \epsilon_a^{(I)}(k)V(k)^{(I)} \ \ \mbox{with} \ \ V(k)^{(I)}=d_{k}^{(I)}\exp{(ik^ax_a)} ,
\eea
where $\epsilon_a^{(I)}(k)$ and $d_k^{(I)}$ with $(I=1,2,3)$ denote three independent  polarization vectors   
and expansion constants respectively.  
And for more simple treatment we choose a co-moving coordinate frame, where 
fields are at rest, that is $k^a=(M, 0, 0, 0)$.  
The supplementary condition eq.(\ref{supplementary}) 
is now $k^a\epsilon_a^{(I)}(k)=0$ and three independent solutions are simply obtained: 
\bea
\epsilon_a^{(1)}=(0,1,0,0) , \ \ \epsilon_a^{(2)}=(0,0,1,0),\ \ \epsilon_a^{(3)}=(0,0,01). 
\eea

The bi-spinor $\Psi_V(x)$ is decomposed into the ket and bra spinors, which are defined: 
\bea
|V(k)^{(I, j)}>&\equiv &(M/2-\gamma^a\partial_a)\sqrt{2V(k)^{(I)}}|j> , \label{defVket} \\
<V(k)^{(I, j)}|&\equiv &<j|\sqrt{2V(k)^{(I)}}(M/2-\overleftarrow{\partial_a}\gamma^a)\gamma^a\epsilon_a^{(I)}.\label{defVbra}
\eea
If we use the explicit form of unit spinors (\ref{unit}), we obtain non vanishing components of ket spinors 
common for $I=1,2,3$ as 
\bea
|V^{(I,j=1)}>=M\sqrt{2V(k)^{(I)}}|\hat{u}_{\uparrow}>,\ \ 
|V^{(I,j=2)}|=M\sqrt{2V(k)^{(I)}}|\hat{u}_{\downarrow}>,
\eea
 where the unit ket spinors are: 
\bea
|\hat{u}_{\uparrow}>=(1,0,0,0)^T,  \ \ |\hat{u}_{\downarrow}>=(0,1,0,0)^T ,  
\eea 
where $T$ denotes the transpose operation. 
And we obtain non vanishing components of bra spinors as
\bea
<V^{(I=1,j=1)}|=iM\sqrt{2V(k)^{(I=1)}}<\hat{v}_{\uparrow}|,\ \ 
<V^{(I=1,j=2)}|=-iM\sqrt{2V(k)^{(I=1)}}<\hat{v}_{\downarrow}|,
\eea
for $\epsilon_a^{(I=1)}$ and
\bea
<V^{(I=2,j=1)}|=M\sqrt{2V(k)^{(I=2)}}<\hat{v}_{\uparrow}|,\ \ 
<V^{(I=2,j=2}|=M\sqrt{2V(k)^{(I=2)}}<\hat{v}_{\uparrow}|,
\eea
for $\epsilon_a^{(I=2)}$ and
\bea
<V^{(I=3,j=1)}|=-iM\sqrt{2V(k)^{(I=3)}}<\hat{v}_{\uparrow}|,\ \ <V^{(I=3,j=2}|=-iM\sqrt{2V(k)^{(I=3)}}<\hat{v}_{\uparrow}|,
\eea
for $\epsilon_a^{(I=3)}$, where the unit bra spinors are: 
\bea
<\hat{v}_{\uparrow}|\equiv (0,0,0,1), \ \ <\hat{v}_{\downarrow}|\equiv (0,0,1,0). 
\eea 
Using these ket and bra spinors, the bi-spinor for the vector fields is reconstructed as 
\bea
\sum_{j=1}^{4}|V(k)^{(I=1,j)}><V(k)^{(I=1,j)}|&=&i2M^2(|\hat{u}_{\uparrow}><\hat{v}_{\uparrow}|-
|\hat{u}_{\downarrow}><\hat{v}_{\downarrow}|), \nn\\
&=&M(M-\gamma^a\partial_a)\gamma^b\epsilon_b^{(I=1)}V(k)^{(I=1)} 
\eea
for  $\epsilon_a^{(I=1)}$ and
\bea
\sum_{j=1}^{4}|V(k)^{(I=2,j)}><V(k)^{(I=2,j)}| &=&2M^2(|\hat{u}_{\uparrow}><\hat{v}_{\uparrow}|+
|\hat{u}_{\downarrow}><\hat{v}_{\downarrow}|), \nn\\
&=&M(M-\gamma^a\partial_a)\gamma^b\epsilon_b^{(I=2)}V(k)^{(I=2)}
\eea
for  $\epsilon_a^{(I=2)}$ and
\bea
\sum_{j=1}^{4}|V(k)^{(I=3,j)}><V(k)^{(I=3,j)}| &=&-i2M^2(|\hat{u}_{\uparrow}><\hat{v}_{\downarrow}|+
|\hat{u}_{\downarrow}><\hat{v}_{\uparrow}|), \nn\\
&=&M(M-\gamma^a\partial_a)\gamma^b\epsilon_b^{(I=3)}V(k)^{(I=3)}
\eea
for  $\epsilon_a^{(I=3)}$. 
These expressions of spin structures in the product of spin 1/2 spinors show that 
they form three components of spin one vector fields. 

%%%%%%%%%%%%%%%%%%%%%%%%%%%%%%%%%%%%%%%%%%%%%%%%%%%%%%%%%%%%%%

\end{document}